\documentclass[preprint,prb,epsf,showpacs,superscriptaddress]{revtex4}
\usepackage{epsf}
\usepackage{graphicx}

\begin{document}

\title{The magnetic susceptibility of exchange-disordered 
antiferromagnetic finite chains}
\author{C.M. \surname {Chaves}}
\email{cmch@if.ufrj.br}
\affiliation{Instituto de F\'\i sica, Universidade Federal do Rio de Janeiro, Caixa
Postal 68528, 21941-972, Rio de Janeiro RJ, Brazil}
\author{ Thereza \surname {Paiva}}
\affiliation{Instituto de F\'\i sica, Universidade Federal do Rio de Janeiro, Caixa
Postal 68528, 21941-972, Rio de Janeiro RJ, Brazil}
\author{ J. d'Albuquerque \surname{e Castro}}
\affiliation{Instituto de F\'\i sica, Universidade Federal do Rio de Janeiro, Caixa
Postal 68528, 21941-972, Rio de Janeiro RJ, Brazil}
\author{ F.~\surname{H\'ebert}}
\affiliation{Institut Non-lin\'eaire de Nice, Universit\'e de Nice-Sophia Antipolis, 1361
route des Lucioles, 06560 Valbonne, France }
\author{ R. T. \surname{Scalettar}}
\affiliation{Department of Physics, University of California, Davis, California 95616}
\author{ G. G. \surname{Batrouni}}
\affiliation{Institut Non-lin\'eaire de Nice, Universit\'e de Nice-Sophia Antipolis, 1361
route des Lucioles, 06560 Valbonne, France }
\author{Belita \surname {Koiller}}
\affiliation{Instituto de F\'\i sica, Universidade Federal do Rio de Janeiro, Caixa
Postal 68528, 21941-972, Rio de Janeiro RJ, Brazil}
\date{\today }

\begin{abstract}
The low-temperature behavior of the static magnetic susceptibility $\chi(T)$ of
exchange-disordered antiferromagnetic spin chains is investigated. It is
shown that for a relatively small and even number of spins in the chain, 
two exchange distributions which are expected to occur in nanochains of P
donors in silicon lead to qualitatively distinct behaviors of the low-temperature 
susceptibility. As a consequence,  magnetic measurements 
might be useful to characterize whether a given sample 
meets the requirements compatible with Kane's original proposal
for the exchange gates in a silicon-based quantum computer hardware. 
We also explore the dependence of $\chi(T)$ on the number of spins in the chain as it 
increases towards the thermodynamic limit, where any degree or 
distribution of disorder leads to the same low-temperature scaling behavior. 
We identify a crossover regime where the two distributions of disorder may not be clearly differentiated, but the characteristic scaling of the thermodynamic limit has not yet been reached.

\end{abstract}

\maketitle
\section{Introduction}
\label{sec:intro}

Low dimensional quantum antiferromagnets have been intensively studied over
the years, and became a reference problem in Condensed Matter Physics,
Strongly Correlated Systems, and Statistical Mechanics.\cite{general-1D} 
Quantum spin chains may exhibit quantum phase transitions
and, in the continuum limit, turned out to be
an active area for application of quantum field theory in Condensed Matter
Physics. Random quantum antiferromagnetic (AF) chains, which model the
magnetic behavior of several quasi-one-dimensional compounds, have also
attracted a great deal of attention\cite{villain}.
Theoretical studies\cite{ma1,daniel}  of the quantum spin-1/2 disordered Heisenberg 
antiferromagnet in 1D show that, for infinite chains, 
the magnetic susceptibility behavior is essentially 
insensitive to the specific disorder distribution.
While the behavior in the thermodynamic limit
is thus well established, there has not yet
been a detailed study of how that limit is approached, and specifically
how the behavior of the magnetic susceptibilities of finite chains
might depend on the disorder distribution.  This is an interesting
question in light of current efforts towards nanoscale control and applications which, coupled to improved techniques of material growth, directed interest into magnetic nanostructures, 
in particular into finite magnetic chains.\cite{Gambardella} 
 
One of the most exciting potential applications of magnetism at the extreme microscopic level 
consists in utilizing the 2-level dynamics of the electron spin as a physical implementation of the quantum-bit (qubit) in a solid state quantum computer, where the required entanglement between qubits could be provided by the exchange coupling between electrons.\cite{LD}    
For example, the quantum behavior of exchange-coupled electrons bound to an array of 
phosphorous donors in silicon is a key element in  
Kane's proposal for a Si-based quantum computer.\cite{Kane} 
Low-temperature magnetic susceptibility measurements of P doped Si have 
provided valuable information regarding the exchange distributions in  
randomly 3D-doped samples.\cite{Andres} 
It is expected that similar measurements might also be relevant in 
characterizing linear arrays of donors.
In Kane's proposal, two-qubit operations are mediated by the exchange 
interaction between electrons bound to nearest-neighbor P atoms in the chain. 
Previous studies have shown that this exchange coupling $J$ 
is always AF, and that its strength is highly sensitive to 
the inter-donor relative positioning.\cite{Andres,bk1,bk2} 
Indeed, changes of just one lattice parameter in the relative 
positioning of two P impurities may alter the
magnitude of the coupling between them by orders of magnitude. 
Controllable exchange coupling compatible 
with Kane's original proposal would be achieved if all donors in the 
chain could be positioned \textit{exactly} along a single [100] 
crystal axis.\cite{bk2} 
In this situation, the exchange coupling behavior is (as assumed by Kane\cite{Kane}) similar to the hydrogenic case,\cite{hf} 
 i.e., it decays exponentially with increasing interdonor distance. 
Uncertainties in the interdonor distances {\it along the chain} would result in 
a narrow distribution of values for $J$ around a ``target'' value $J_0$.
Assuming a substitutional donor positioning precision of about 1 nm in the 
Si lattice,\cite{fabrication1} the perfect [100] alignment situation may be modeled by a trimodal exchange distribution $P_{tri}(J)$. 
However, if instead of perfect alignment the donor positions 
are randomly distributed among all substitutional sites within a small spherical region of 1 nm around the ideal (``target'') impurity sites, 
the peculiar band structure of Si leads to a wide distribution of  
exchange coupling, peaked at $J=0$,\cite{bk2} 
causing difficulties in the operation and control of the ``exchange gates''.
We approximate such distribution here by an exponential function, 
$P_{exp}(J)$ for $J>0$.  
It is therefore clear that the exchange disorder distribution is also highly sensitive to the 
positioning distribution of the P donors in Si.

The magnetic susceptibility behavior of 
{\em infinite chains} is essentially insensitive to the 
specific disorder distribution, as mentioned above, indicating that 
susceptibility measurements would not differentiate among the two cases 
mentioned above in this limit. Of course this may be different 
for {\em finite chains}, a situation which is also of practical interest in terms of 
guiding current fabrication efforts towards P donors positioning in Si.\cite{fabrication1,fabrication2} The aim of the present work is to shed light on this problem 
by investigating the relation between the magnetic response of linear chains 
of spins and the distributions of the exchange interaction within the chains.
Our results indicate that for AF disordered chains with even and 
relatively small number $N$ of sites, the exchange distributions 
$P_{tri}(J)$ and $P_{exp }(J)$
lead to quite distinct low temperature behavior of the zero-frequency uniform
magnetic susceptibility. 
Hence, the two dis\-tri\-bu\-tions could be experimentally distinguished 
by sufficiently sensitive magnetic measurements, providing useful information 
regarding donor alignment.

This paper is organized as follows. In Sec.~\ref{sec:infi} we briefly review results available in the literature regarding infinite AF chains. In Sec.~\ref{sec:finite} we consider finite chains with relatively small number of spins, starting from spin pairs and trios for which analytical solutions are obtained, as well as 8-spin chains, which are solved numerically. In Sec.~\ref{sec:cross} we analyze the crossover into the thermodynamic limit by solving for longer chains via a Quantum Monte-Carlo method. Our summary and conclusions are presented in Sec.~\ref{sec:conclude}.

\section {Infinite Antiferromagnetic Chains}
\label{sec:infi}

The Hamiltonian describing an open chain with $N$ spins is
\begin{equation}
\mathcal{H}=\sum_{i}^{N-1}J_{i}\vec{S_{i}}\cdot \vec{S}_{i+1},  \label{H}
\end{equation}
where the spin quantum number is $S=1/2$. 
Since we are interested in AF chains, we assume that $J_{i}\geq 0$ for every $i$.
We briefly review in this section several pertinent results available in the
literature for AF chains in the $N\to\infty$ limit.  

The ground state of an {\it ordered} infinite chain $(J_{i}=J$ for every $i$) can be obtained
from Bethe ansatz. Griffiths\cite{griffiths} has shown that the zero
temperature susceptibility per spin of such system is finite and given by $%
\chi (T=0)/\chi _{0}(J)=1/\pi ^{2},$ where $\chi _{0}(J)=g^{2}(\mu _{B})^{2}/J$, where $\mu _{B}$ is the Bohr magneton and $g(=2)$ is the Land\'{e} factor.
For general $T$, field theory methods [$k=1$ Wess-Zumino-Witten (WZW)
non-linear $\sigma $ model] \cite{affleck} give%
\begin{equation}
\frac{\chi (T)}{\chi _{0}(J)}=\frac{1}{(\pi )^{2}}\left( 1+\frac{1}{2\ln(T_{0}/T)}%
\right) ,  \label{affleck}
\end{equation}%
where $T_{0}$ is a temperature cutoff. Quantum Monte-Carlo (QMC) calculations of $%
\chi (T)$ have been carried out by Kim \textit{et al.}\cite{Kim} and their
results for $T_{0}/J=1.8$ (we use units of energy for temperature, that is, the Boltzmann constant $k_{B}$ is set equal to one) are well fitted by the WZW expression (\ref{affleck}). 

According to real-space renormalization group theory,\cite{daniel}
the introduction of any amount of disorder drives the system into a random 
singlet phase, in which each spin forms a singlet pair with another spin; pairs with 
arbitrarily long distance also exist. 
Bonds among distant spins, however, correspond to very weak 
coupling. The low-temperature excitations basically involve breaking 
these weakest bonds, resulting in nearly-free spins giving rise to a Curie 
susceptibility modified by the statistics in the number of contributing spins.
As a result the magnetic susceptibility at low $T$ diverges as\cite{daniel} 
\begin{equation}
\chi(T\to 0) \sim 1/\left[T(\log T)^{2}\right].
\label{lt}
\end{equation}

\section {Finite Chains}
\label{sec:finite}

The above results refer to chains in the thermodynamic limit ($%
N\rightarrow \infty $). For \textit{finite ordered} rings (periodic
boundary conditions), Bonner and Fisher \cite{bonner} have calculated
the susceptibility per spin $\chi_N (T)$ for up to N=11 spins based on direct  
diagonalization of $\mathcal{H}$ in the presence of a magnetic field. 
They have found that $\chi_N (T\to 0)$ exhibits distinct behavior 
depending on whether $N$ is even or odd. In the first
case, pairs of neighboring spins tend to form singlets and 
$\chi_{N=even}(T\to 0) \rightarrow 0$, whereas in the second case, 
the occurrence of  unpaired spins leads to 
a Curie-law behavior 
$\chi_{N=odd} (T \to 0) \sim 1/T \rightarrow \infty $. 
These results immediately raise the question as
to what extent the behavior of $\chi_N(T) $ changes by the
introduction of disorder. Moreover, could we infer, based on the magnetic
response, the type of exchange disorder distribution?
The relevant distributions here are: (i) trimodal, with 
\begin{equation}
P_{tri}(J)=(1/3) \{ \delta (J-J_{0})+
\delta (J-(1+W)J_{0})+\delta (J-(1-W)J_{0})\}~, 
\end{equation}
where $J_{0}$ and $W$ are both positive, with $W<1$, 
and (ii) exponential, with 
\begin{equation}
P_{exp }(J)=\frac{1}{J_{0}}e^{-J/J_{0}} \, \Theta(J) ~,
\end{equation}
where $\Theta$ is the step-function. 
Note that in both cases $\langle J \rangle = J_0$, the exchange ``target'' value.

We consider initially the $N=2$ case, for which the
susceptibility per spin is given by\cite{haraldsen} 
\begin{equation}
\frac{\chi _{2}(T,J)}{\chi_{0}(J)}=\frac{\beta J}{3+e^{\beta J}},  \label{chi2}
\end{equation}
where $\beta =1/T$ and $\chi_0(J)$ is given above Eq.~(\ref{affleck}). 
Considering the average of $\chi _{2}$ over the above
distributions, it is
clear that 
$\left\langle \chi _{2}\right\rangle _{tri}=\int\limits_{0}^{\infty
}P_{tri}(J)\ \chi _{2}(T,J)\ dJ\   \label{chi2tri}$
vanishes as $T\rightarrow 0$, as in the absence of disorder. 
For the exponential distribution,  it is convenient to 
split the integral for $\left\langle \chi_{2}\right\rangle_{exp }$ 
into two terms
\begin{equation}
\langle \chi _{2}\rangle _{exp }=\frac{\beta}{J_0} \int_{0}^{\alpha /\beta }
\frac{e^{-J/J_0}%
}{3+e^{\beta J}}\,dJ+\frac{\beta}{J_0} \int_{\alpha /\beta }^{\infty }
\frac{e^{-J/J_0}}{%
3+e^{\beta J}}\,dJ,  \label{chi2e}
\end{equation}%
where $\alpha $ is a constant,  
and $\langle \chi _{2}\rangle _{exp }$ is given in units of $(g\mu
_{B})^{2}$. For sufficiently low temperatures, $\alpha $ can be chosen such
that $\alpha /\beta \ll J_0$ and $e^{-\alpha }<<1$. Hence, in the first
integral (corresponding to small values of $J$), we can approximate $%
e^{-J/J_0}\approx 1$, while in the second one, the term $e^{\beta J}$ is always
much greater than 1, leading to 
\begin{equation}
J_0\langle \chi _{2}\rangle _{exp }\approx C_{1}+e^{-\alpha }\approx C_{1},  \label{chi2eT0}
\end{equation}%
where $C_{1}=\int_{0}^{\alpha }1/(3+e^{x})\ dx\approx $ $\int_{0}^{\infty
}1/(3+e^{x})\ dx = \ln(4)/3$. Thus, as $T\rightarrow 0$, 
$\langle \chi _{2}\rangle _{exp }$ approaches a
non-zero value, in contrast with the trimodal distribution result.

For an open $N=3$ chain, the susceptibility per spin is\cite{haraldsen}
\begin{equation}
\chi _{3}(T,J_{1},J_{2})=(\beta /12)\frac{5+e^{\beta (J_{1}+J_{2})/2}\cosh
\left( (\beta/2) \sqrt{J_{1}^{2}-J_{1}J_{2}+J_{2}^{2}}\right) }{1+e^{\beta
(J_{1}+J_{2})/2}\cosh \left( (\beta/2) \sqrt{J_{1}^{2}-J_{1}J_{2}+J_{2}^{2}}%
\right) }.  \label{chi3}
\end{equation}
It is clear that for non-negative values of $J_{1}$ and $J_{2}$, $\chi_{3}$
exhibits a Curie-like divergence as $T\rightarrow 0$. As a consequence, the
average of $\chi _{3}$ over the trimodal distribution, $\left\langle \chi
_{3}\right\rangle _{tri}$, also diverges as $T$ approaches 0. 
Regarding the exponential distribution, 
the change of variables $J_{1}=J\sin \theta $ and $J_{2}=J\cos \theta $ 
leads to 
%
\begin{equation}
J_0^2\langle \chi _{3}\rangle _{exp }=(\beta /12)\int_{0}^{\pi /2}d\theta
\int_{0}^{\infty }\,e^{-(\sin \theta +\cos \theta )J/J_0}\frac{5+f(\beta
J,\theta )}{1+f(\beta J,\theta )}\ J\ dJ,  
\label{chi3exp}
\end{equation}
where $f(x,\theta )=e^{x(\sin \theta +\cos \theta )/2}\cosh
\left( (x/2)\sqrt{1-\sin \theta \cos \theta }\right) $. As for the $N=2$
case, we split the integral above into
two, which we label $I_{1}$ and $I_{2}$, corresponding to $0\leq J\leq
\alpha /\beta $ and $\alpha /\beta \leq J\leq \infty $, respectively. 
For sufficiently low $T$, $\alpha $ is again chosen such that $\alpha
/\beta \ll J_0$ and $e^{-\alpha }<<1$. Since $1\leq \sin \theta +\cos
\theta \leq \sqrt{2}$, the term $e^{-(\sin \theta +\cos \theta )J/J_0}$ in the
first integral (corresponding to small values of $J$) can be approximated by
1, giving 
\begin{equation}
I_{1}\approx \frac{1}{\beta ^{2}}\int_{0}^{\pi /2}d\theta \int_{0}^{\alpha}\,%
\frac{5+f(x,\theta )}{1+f(x,\theta )}\ x\ dx=C_{1}^{\prime }/\beta ^{2},
\label{i1}
\end{equation}%
where $C_{1}^{\prime }$ is $T$-independent. In the second integral, since $\beta J\geq
\alpha $ and $\alpha $ is large, the $T$-dependent term in both the
numerator and denominator of the integrand is much larger than 1, so that
\begin{equation}
I_{2}\approx \int_{0}^{\pi /2}d\theta \int_{\alpha /\beta }^{\infty
}\,e^{-(\sin \theta +\cos \theta )J/J_0}J\ dJ\approx \int_{0}^{\pi /2}d\theta
\int_{0}^{\infty }\,e^{-(\sin \theta +\cos \theta )J/J_0}J\ dJ=C_{2}^{\prime },
\label{i2}
\end{equation}%
where $C_{2}^{\prime }$ is $T$-independent. 
Thus, from Eq.~(\ref{chi3exp}), 
$J_0\langle \chi _{3}\rangle_{exp}=(\beta /12)(I_{1}+I_{2})/J_0\approx \beta C_{2}^{\prime }/J_0$,  
which diverges following a Curie law as $T\rightarrow 0$.

For a better physical insight, it is instructive to compare the $N=2$ and $N=3$ cases. 
We note that in the low-$T$ regime, the $T$-dependent behavior of $\chi_N$ 
is dominated in the first case by the disorder distribution , 
while in the second by the odd parity of $N$. 
The Curie-type low-$T$ behavior of $\langle\chi_3(T)\rangle$ is related to
the occurrence of  ``unpaired'' spins, independently of  $P(J)$. 
It is interesting that the large-$J$ tail of the 
distribution gives the dominant contribution to the average behavior, namely the 
integral given by $I_2$ in Eq.~(\ref{i2}), corresponding to the 
more strongly coupled 3-spin chains.  
On the other hand, for $N=2$, the behavior of $\langle\chi_2(T)\rangle$ at low $T$ 
is dominated by the contribution from small values of $J$ (more
precisely, from values of $J\ll min\{T,J_{0}\}$), as  
shown in Eqs.~(\ref{chi2e}) and (\ref{chi2eT0}). 
The occurrence of arbitrarily small values of $J$ in the exponential distribution weakens the
tendency of neighboring spins to form singlets,\cite{bonner} leading to a finite value 
of $\langle\chi_2(T=0)\rangle_{exp}$, while $\langle\chi_2(T=0)\rangle_{tri}=0$.

\begin{figure}

\resizebox{160mm}{!}
{\includegraphics{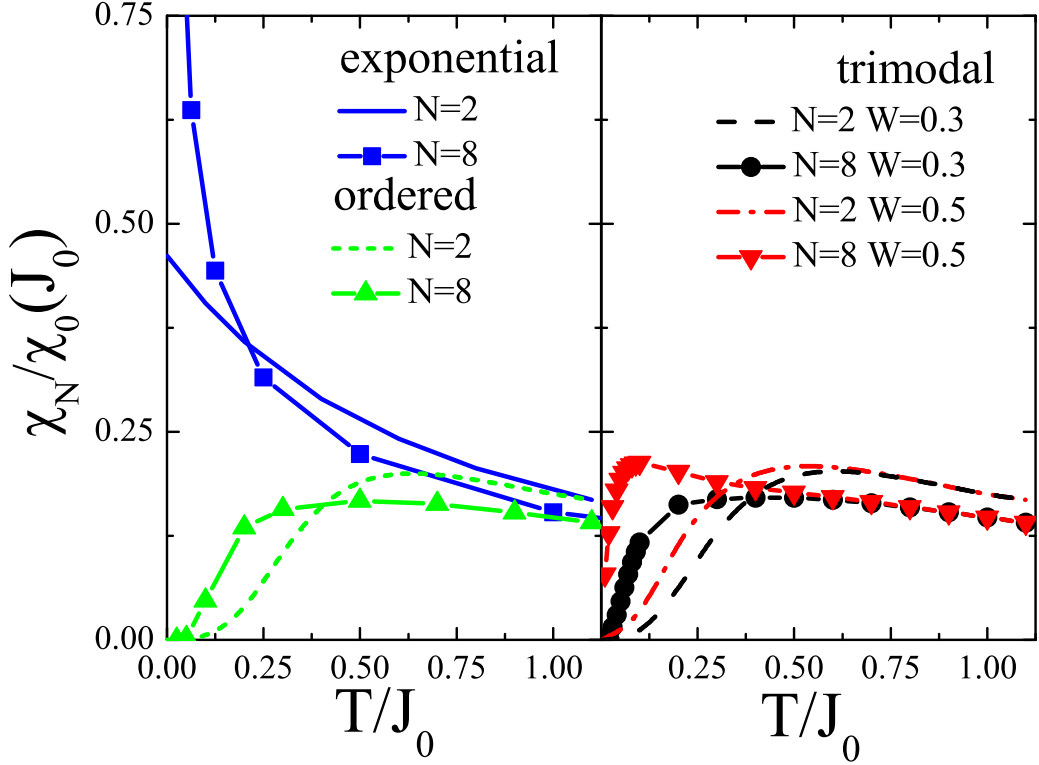}}
\caption{Low-temperature behavior of the magnetic susceptibility per spin of disordered 
antiferromagnetic spin chains with exponential and trimodal distributions of exchange coupling. 
The  plots give $\chi_N/\chi_0(J_0)$ versus $T/J_0$ for $N=2$ and $8$.
Results for the exponential disorder distribution and for ordered chains are given on the left panel, 
whereas those for the trimodal disorder distributions with width $W=0.3$ and $0.5$ are given on the right panel. 
All results are for open boundary conditions. 
For $N=8$ the statistical error bars are smaller than the data points, and the straight lines 
joining the points are just guides to the eye. 
}
\label{f2}
\end{figure}

We expect the above considerations to apply to other finite values of $N$.
As odd-parity $N$ chains always lead to unpaired spins, and thus to a 
Curie-like low-$T$ divergence of $\langle\chi_{N=odd}(T)\rangle$,
susceptibility measurements would not be useful for distinguishing among 
different exchange disorder distributions if $N$ is odd.  
Since analytical calculations become impractical as $N$ increases, 
in order to illustrate the even-parity case 
we have carried out numerical calculations for both $\langle\chi_8\rangle_{tri}$
and $\langle\chi_8\rangle_{exp}$ as a function of $T$.
We obtain the susceptibility numerically by 
determining the spin-spin correlation function $\langle S^{z} _i S^{z}_j\rangle$ from which, through the fluctuation-dissipation theorem, we obtain
\text{$\chi = \beta \sum_{i,j}\langle S^{z} _i S^{z}_j\rangle$}.  
For each temperature we have averaged over 10,000 realizations of disordered configurations. 
For the trimodal distribution, two values for the width parameter have been 
considered, namely $W=0.3$ and $0.5$. 
Results are presented in Fig.~\ref{f2}, which shows susceptibility curves for 
chains with $N=2$ and 8, for trimodal and exponential disorder distributions, as well as for the 
ordered chain cases. 
We note that results for $\langle\chi_{N}(T)\rangle_{tri}$ are qualitatively very similar, regardless of the width parameter $W$, even in the limit $W=0$, corresponding to the ordered chains. 
All curves reach a maximum and eventually decrease as $T \to 0$, going to zero for $T=0$. 
We remark the obvious fact, also illustrated in Fig.~\ref{f2}, that as $W$ increases the maximum in $\chi$ and the sharp downturn toward zero value occur at lower temperatures.
The results for the exponential distribution are markedly different, with an increasing 
$\langle\chi _{N}(T)\rangle_{exp }$ for decreasing $T$. 
In principle the trimodal and exponential distributions might be identified and differentiated 
through low-temperature susceptibility measurements in such even-parity chains.

\section{Crossover regime and Thermodynamic Limit}
\label{sec:cross}

Further increase in $N$ eventually leads to the thermodynamic limit behavior\cite{todo}  given in Eq.~(\ref{lt}).
For even $N$, approach to such behavior, giving a divergent $\chi$ as $T\to 0$, might be expected from the comparison between the results of $\langle\chi _{2}(T)\rangle_{exp}$ and $\langle\chi _{8}(T)\rangle_{exp}$ in Fig.~\ref{f2}, but it is not so clear for the trimodal distributions. Another puzzling point regards the sensitivity to even $(N_{even})$ or odd $(N_{odd})$  values of $N$. For small-$N_{odd}$, $\chi$  exhibits a Curie-like $\sim 1/T$ divergence at $T\to 0$ independent of the disorder distribution, as discussed for $N=3$ in Sec.~\ref{sec:finite}, while small-$N_{even}$ chains are quite sensitive to $P(J)$. But odd- and even-chains results must approach each other, possibly with some sensitivity to the type of disorder (remanent from $N_{even}$), as $N$ increases.

In order to clarify these points we have calculated the susceptibility for larger values of $N$ using a stochastic series expansion (SSE) method. The SSE is a QMC method based on the Taylor expansion of the Boltzmann weight operator $e^{-\beta \mathcal{H}}$ up to a very high order.\cite{anders} The partition function and observables can then be evaluated via importance sampling of the different terms appearing in this series.
Choosing a large enough order for the expansion, the systematic errors introduced by the truncation of the series are negligible. The SSE method also allows us to use importance sampling update schemes based on global changes of the system (cluster or loop updates) that are extremely efficient, especially for the highly symmetric Heisenberg model studied here.
We have adjusted the precision of the data obtained through importance sampling so that the errors obtained on each individual realization are roughly one order if magnitude smaller than a typical difference between two realizations. We have considered open chains and averaged our results over 10,000 disorder realizations (except for the large size $N=128$ where we used  periodic boundary conditions and averaged over only 100 realizations).

\begin{figure}
\resizebox{180mm}{!}
{\includegraphics{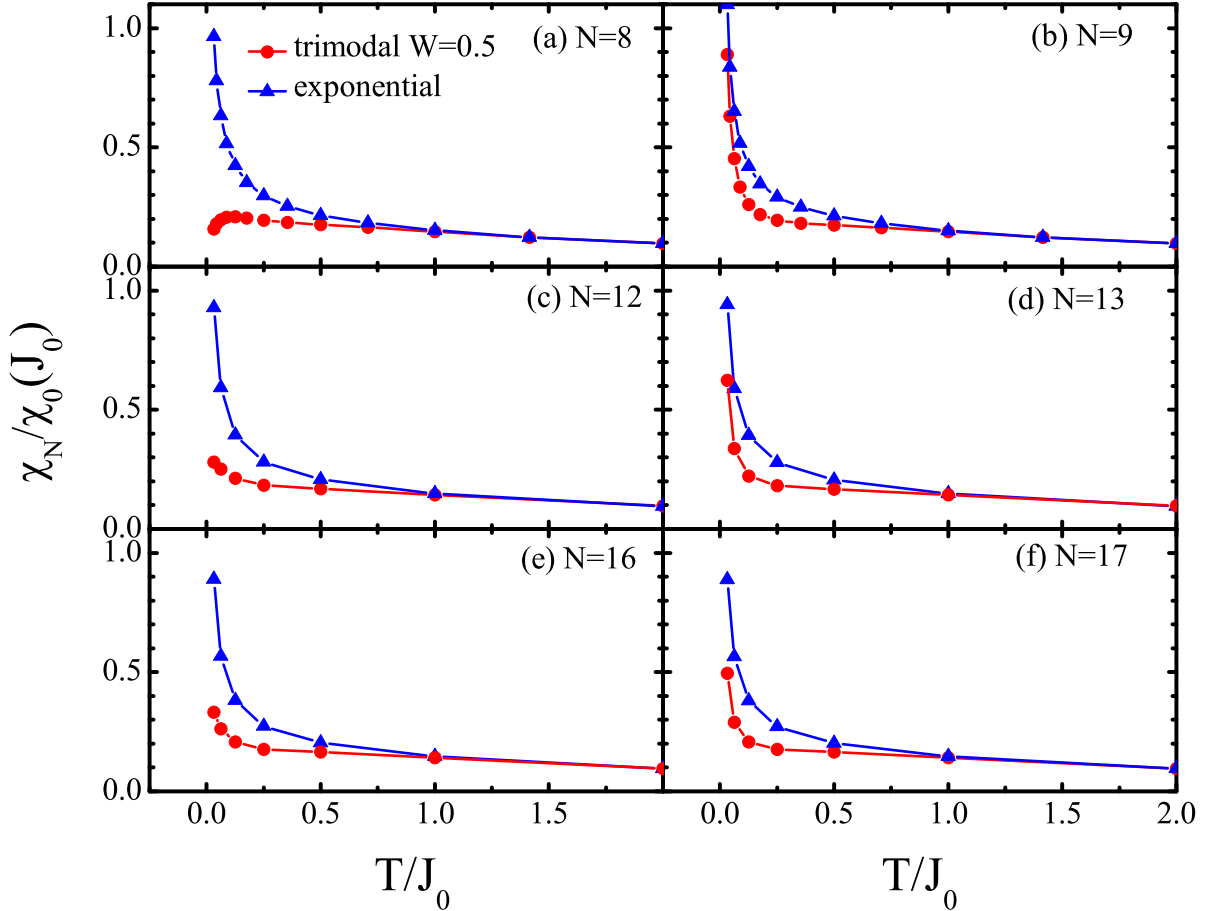}}
\caption{Magnetic susceptibility per spin for disordered finite chains of $N$ spins for trimodal  $(W=0.5)$  and exponential disorder distributions. For $T\to 0$ the $N=8$ results for the  two distributions are clearly distinct, while for $N=9$ they are practically collapsed.
As the number of spins in the chain increases, the results for the two distributions and $N_{even}$ (left column) approach each other,  as results for $N_{even}$ and $N_{even}+1$ chains (successive rows) also approach each other. Lines are guides to the eye and statistical error bars are smaller than the data points.}
\label{fLT}
\end{figure}

\begin{figure}
\resizebox{180mm}{!}
{\includegraphics{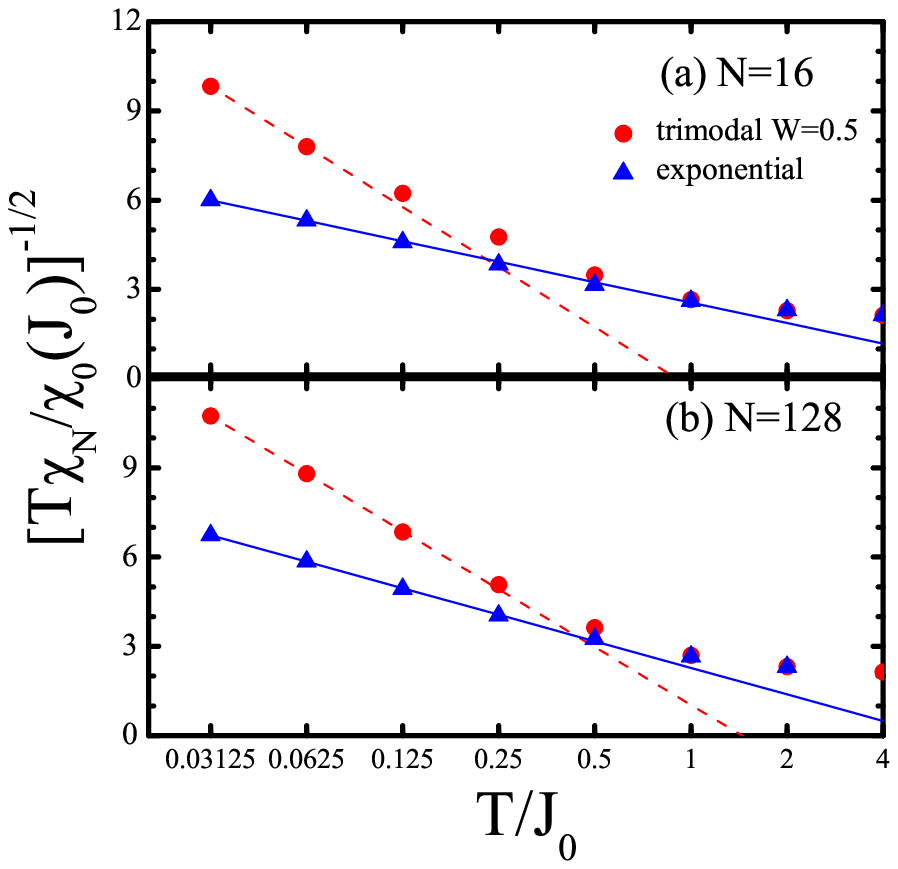}}
\caption {Scaled plots for the low-temperature behavior of the magnetic susceptibility per spin of disordered chains, illustrating the approach to the thermodynamic limit as $N$ increases.
The plots give $1 /\sqrt{T\chi_N/\chi_0(J_0)}$ versus $T/J_0$ (logarithm scale) for exponential and trimodal  $(W=0.5)$ distributions in chains with $N=16$ and 128. The thermodynamic limit $(N\to \infty)$ corresponds to the linear behavior indicated by the straight solid and dashed lines drawn to connect the 
two lowest-$T$ data points for exponential and trimodal distributions 
respectively.} 
\label{f3}
\end{figure}

We present in Fig.~\ref{fLT} results for $\langle\chi_N (T) \rangle_{tri}$ and $\langle\chi_N (T)\rangle_{exp}$ for increasing $N$ up to $N=17$ for the exponential and trimodal $(W=0.5)$ distributions. 
We identify here the following trends towards the thermodynamic limit: (i) from the frames for $N_{even}$ (on the left) we see that the well differentiated disorder distribution results for $N=8$ in (a) approach each other as $N_{even}$ increases [(c) and (e)]; (ii) from the frames for $N_{odd}$ (on the right) we do not identify significative changes as the Curie behavior discussed in Sec.~\ref{sec:finite} for N=3 acquires logarithmic corrections [eventually leading to the asymptotic behavior governed by Eq.~(\ref{lt})] which are not easily captured on this scale and through this range of $N_{odd}$ values; (iii) following successive rows we see that the even-odd differences (namely going from $N_{even}$ to $N_{even}+1$) become less prominent as $N_{even}$ increases [(a)-(b), (c)-(d), (e)-(f)]. 
From the practical point of view, it is clear from (i) that, contrary to the $N=8$ results, for $N = 16$ the two distributions lead to very similar qualitative types of behavior of the susceptibility, indicating that, for the particular distributions and temperature range considered here, they may not be clearly differentiated beyond this size of chains, regardless of the parity of $N$.

Approach to the thermodynamic limit, given in Eq.~(\ref{lt}), is usually investigated by plotting $(T\chi)^{-1/2}$ vs $\log T/J_0$, which leads to a linear behavior at low temperatures in this limit.\cite{todo} 
In Fig.~\ref{f3} we illustrate the crossover regime by presenting such scaled plots for $N=16$ and 128 for trimodal $(W=0.5)$ and exponential distributions. In all cases, straight lines (dashed for trimodal and solid for exponential) are drawn through the two lowest-$T$ calculated points. 
We note that  for the trimodal distribution and $N=16$ such line does not include any other calculated point, whereas for $N=128$ it gives a good fitting for $T/J_0$ up to about $ 0.25$, indicating  that the thermodynamic limit has been reached up to this temperature for this value of $N$. 
For the exponential distribution, the approach to the thermodynamic limit scaling with increasing $N$ is faster, although careful analysis of the data plotted in Fig.~\ref{f3} shows that the points for $N=16$  and exponential distribution [triangles in (a)] down-turn from the solid line at small $T$, and the good agreement for $T/J_0=1$ in (a) is fortuitous, as it does not remain for $N=128$ [triangles in (b)]. 
It is interesting to note that the solid (dashed) line in Fig.~\ref{f3}(a) is nearly parallel to the solid (dashed) line in (b), indicating that some aspects of the $N\to \infty$ behavior are already captured at the smaller $N$ values at low-$T$. 
Fig.~\ref{f3}(b) shows that in the thermodynamic limit the susceptibility is quantitatively quite sensitive to the disorder distribution, although in practice this is probably not as valuable a tool to identify the disorder distribution as the differences encountered for the smaller even values of $N$ [e.g. Fig.~\ref{fLT}(a)]. 

\section{Summary and conclusions}
\label{sec:conclude}

We have investigated the low-$T$ behavior of the uniform magnetic susceptibility of exchange disordered spin-1/2  AF chains as a function of the number $N$ of spins in the chain and disorder, for trimodal and exponential disorder distributions. Formally, the key distinction among the two distributions considered here is that the exponential distribution is not bound, so that pair exchange coupling $J$ arbitrarily close to zero may occur, while the trimodal distribution is bound and $J$ does not become arbitrarily small.  

For chains with even and relatively small number of spins ($N_{even}\lesssim 8$) the susceptibility displays distinct behaviors for the two exchange distributions which are expected to occur in nanochains of  P donors in Si. According to our results in Fig.~\ref{f2}, it might be possible to identify the atomic-scale positioning of the P atoms in such chains, thus providing complementary information on whether sample preparation techniques meet the requirements compatible with Kane's original proposal for a quantum computer hardware. For larger values of $N$ ($N\gtrsim 16$) such differentiation would not be so straightforward, as illustrated in Fig.~\ref{fLT}. We also note that $\chi_N (T)$ becomes less sensitive to $N$ being even or odd when $N\gtrsim 16$. For completeness, we have also investigated  the approach to the thermodynamic limit, and we have found that for $N=128$ the expected scaling of $\chi(T \to 0)$ is already obtained, while for $N = 16$  the system is still far from the thermodynamic limit, as discussed in Fig.~\ref{f3}. 

We therefore identify three regimes as $N$ increases: (i) when $N\lesssim 8$ the even-$N$ disordered chains present quite distinct behaviors according to the exchange distribution; (ii) for intermediate  $N$  values, illustrated here by $N=16$, the low temperature behavior of the magnetic susceptibility corresponding to different distributions is not easily differen\-tiated, although the thermodynamic limit has not yet been reached; (iii) for larger $N$, illustrated here by $N=128$, the two distributions follow the thermodynamic limit scaling, which is a signature that the random singlet phase is formed at the lowest $T$. 

\begin{acknowledgments}
We thank S.L.A. de Queiroz, R.R. dos Santos and D.J. Priour for helpful suggestions. BK thanks the hospitality of the CMTC at the University of Maryland.
This work has been partially supported in Brazil by  CNPq, FAPERJ, 
the Millennium Institute of Nanoscience and FUJB. RTS acknowledges support from NSF-DMR-0312261 and NSF-INT-0203837.

\end{acknowledgments}

\end{document}